\documentclass[aps,prl,twocolumn,superscriptaddress,floatfix]{revtex4}

\usepackage{amsmath,amssymb}
\usepackage{graphicx,multirow}

\begin{document}

\title{Anomalous Optical Phonon Splittings in Sliding Bilayer Graphene}

\author{Seon-Myeong Choi}
\affiliation
{Korea Institute for Advanced Study, Seoul 130-722, Korea.}
\affiliation
{Department of Physics and Division of Advanced Materials Science,
Pohang University of Science and Technology, Pohang 790-784, Korea.}
\author{Seung-Hoon Jhi}
\email{jhish@postech.ac.kr}
\affiliation
{Department of Physics and Division of Advanced Materials Science,
Pohang University of Science and Technology, Pohang 790-784, Korea.}
\author{Young-Woo Son}
\email{hand@kias.re.kr}
 \affiliation
{Korea Institute for Advanced Study, Seoul 130-722, Korea.}

\begin{abstract}
We study the variations of electron-phonon coupling
and their spectroscopic consequences in response to sliding of two layers
in bilayer graphene using first-principles calculations and a model Hamiltonian.
Our study shows that the long wave-length optical phonon modes
change in a sensitive and unusual way depending on the symmetry as well as
the parity of sliding atomic structures and that, accordingly,
Raman- and infrared-active optical phonon modes behave differently upon the direction
and size of the sliding.
The renormalization of phonon modes by the interlayer electronic
coupling is shown to be crucial to explain their anomalous behavior upon the sliding.
Also, we show that the crystal symmetry change due to the sliding affects
the polarized Stokes Raman-scattering intensity, which can be utilized
to detect tiny misalignment of graphene layers using spectroscopic tools.
\end{abstract}

\maketitle

Successful isolation of graphene~\cite{Novoselov2004} and subsequent experiments that reveal
its special properties~\cite{Novoselov2005,Zhang2005} have generated excitement
to explore the novel properties of the two dimensional (2D) crystal from various disciplines.
Advances in synthesis and experimental techniques
enable the finding of other 2D crystals~\cite{Novoselov2005b} and
the artificial fabrication of multiply-stacked structures~\cite{Nobel}.
Often, the stacking structures lead to very unusual electronic properties
different from those of constituent 2D crystals depending on how they are piled up.
Among them, bilayer graphene (BLG), which is a stacked structure
of two single-layer-graphene (SLG) sheets,
is unique in electronic structure and exhibits colorful variation in low energy
states when its layer stacking is changed.
The in-plane three-fold rotational and mirror symmetries
and the decoupling of strong $\sigma$ and weak $\pi$ bondings
of carbon atoms authorize the uniqueness of graphene systems.
As such, the interlayer coupling in BLG, while a weak van der Waals type, produces
interesting variations in low energy band structures upon changes of stacking geomerties.
For example, its low-energy states in the Bernal stacked pristine form~\cite{McCann}
has quadratic energy bands but change to have linear bands
when it has a rotational staking fault~\cite{Santos,hass100,Shallcross}.
And, a sensitive electronic topological transition is also predicted
for sliding systems~\cite{son84,Falko,Gail}.
Interplay of the interlayer interaction, the electron-phonon coupling,
and the stacking fault by layer slidings
are thus expected to produce drastic changes
in its low energy properties.
Control of its electronic property is also enabled by manipulating such changes.

Recently, an epitaxially grown BLG on the vicinal surfaces
of silicon carbide shows misalignment between two graphene layers exhibiting
a complex nature of its electronic structures~\cite{Hicks}.
Moroever, recent experiments on CVD bi- and trilayer graphene
also reveal possible misaligned layers
at the domain walls in between two ideally stacked graphene systems~\cite{JPark,Hattendorf}.
Immediate questions are how such structural variations are reflected in spectroscopic features
and whether the misalignment of layers can be detected or not.
Raman and infrared (IR) spectroscopies have proven
to be powerful non-destructive methods to study physical and chemical properties of
2D crystals~\cite{dresselhaus,ferrari97,kuzmenko103}.
Physical properties of graphene
under various conditions such as doping or mechanical strains
have been verified using such tools~\cite{dresselhaus,ferrari97,kuzmenko103,ferrari2009,Cheong}.
In consideration of the rapid progress of research
in this field~\cite{son84,Falko,Gail,Hicks,JPark,Hattendorf},
comprehensive analysis of the spectroscopic features of sliding BLG
will provide key information on the stacking geometry, the electron-phonon interactions,
and low energy excitations.

In this paper, we study long wave-length optical phonon modes responsible
for Raman and IR spectrum when two layers of BLG slide each other.
We investigate how the tiny atomic misalignment
between two layers changes its phonon frequencies
and spectroscopic spectrum using first-principles calculations.
It is found that the degeneracy of IR-active modes is immediately lifted by the sliding
and each mode changes separately depending on sliding geometries.
Unlike the behavior of IR modes,
the frequencies of doubly degenerate Raman-active optical phonons
do not change at all in all sliding circumstances.
Such unconventional optical phonon splittings
originate from the difference in their frequency renormalizations
due to interlayer couplings, which is confirmed
by our effective model Hamiltonian calculations.
Moreover, we demonstrate that, owing to its changes in the crystal-symmetry,
non-resonant Stokes Raman-scattering intensities
associated with Raman-active phonons exhibit a strong polarization dependence
so that they can be used to detect sub-Angstrom misalignment
between two layers in sliding BLG.

\section{Results and discussion}

Figure 1 illustrates the sliding geometry of graphene layers in BLG.
When the sliding vector $\vec d$ is along $-\vec\delta_1$,
the BLG gradually transforms
from its pristine form of AB-stacking (designated as AB-BLG)
to AA-stacking ($\vec d=-\vec\delta_1$) in which all carbon
atoms are right on top of each other (AA-BLG).
For the sliding along $+\vec\delta_1$,
the AB-BLG becomes upside-down AB-stacked BLG (BA-BLG).
Equivalent layer stackings are arranged by other sliding vectors
along either $\pm\vec\delta_2$ or $\pm\vec\delta_3$ direction.
Also combination of the sliding vectors is possible as drawn in Fig. 1(a).

\begin{figure}[t]
\includegraphics[width=1.0\columnwidth]{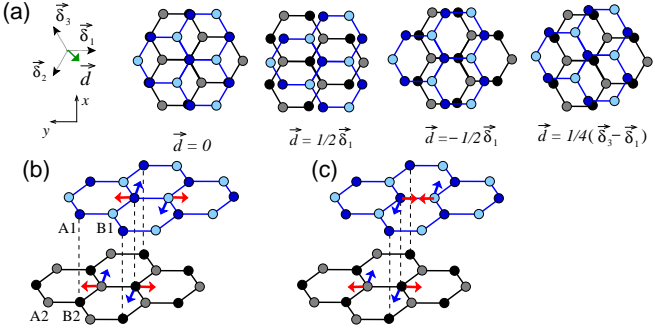}
\caption{(color online)
(a)
$\vec\delta_i (i=1,2,3)$ are
vectors connecting the nearest neighbor carbon atoms
in each layer.
The sliding vector in the upper layer is denoted by $\vec d$.
From the left to the right, top views of BLG for $\vec d=0$, $\frac{1}{2}\vec\delta_1$,
$-\frac{1}{2}\vec\delta_1$, and $\frac{1}{4}(\vec\delta_1 -\vec\delta_2)$.
(b) and (c), in-plane atomic displacements for $E_g$ and $E_u$ optical phonons, respectively.
For each phonon mode, doubly-degenerate atomic motions
are denoted by red (LO-mode)
or blue (TO-mode) arrows with carbon atoms of the same color moving in the same direction.
Vertical dashed-lines are guides for eyes to the nearest interlayer carbon atoms (A2 and B1).
}
\end{figure}

Among the phonon modes at $\Gamma$-point in the Brillouin zone (BZ) of AB-BLG,
high-frequency E$_g$ [Fig. 1(b)] and E$_u$ optical modes [Fig. 1(c)] are
responsible for the Raman G-band
and IR-peak, respectively~\cite{ferrari97,kuzmenko103}.
The in-plane atomic motions in the upper and lower layers
[Figs. 1(b) and (c)] are in-phase and
out-of-phase for E$_g$ and E$_u$ mode, respectively.
These modes are doubly degenerate as denoted
by thick red arrows for longitudinal optical (LO) mode
and by blue ones for transverse optical (TO) mode
as shown in Figs. 1(b) and (c).
Our calculated phonon frequencies of E$_u$ and E$_g$
modes in AB-BLG are 1592 cm$^{-1}$ and 1585 cm$^{-1}$, respectively
(Fig. 2), in good agreement with previous studies~\cite{marzari,tan,Chou}.
As these two modes involve with the interlayer interaction as well as
the crystal symmetry, the sliding motion of two graphene
will imprint its effect on these modes.

\begin{figure}[t]
\includegraphics[width=1.0\columnwidth]{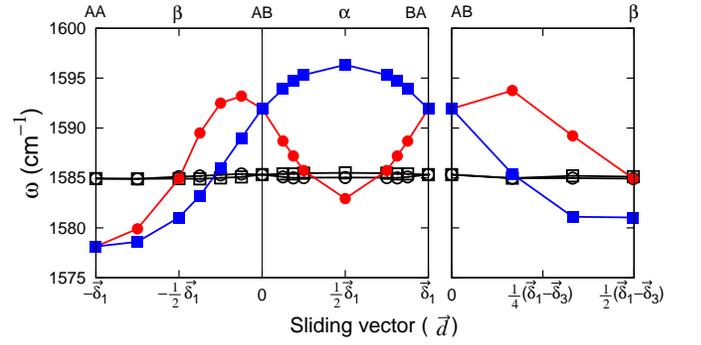}
\caption{(color online)
Variation of optical phonon frequencies as a function of sliding distance and direction.
Red filled circles (blue filled rectangles) are frequencies for IR-active LO (TO) modes
and black empty circles and empty rectangles denote two Raman-active modes.
The sliding vectors are shown in the bottom abscissa and corresponding stacking geometries in the top.
The $\alpha$- ($\vec d = \frac{1}{2}\vec\delta_1$) and $\beta$-stacking
[$\vec d=-\frac{1}{2}\vec\delta_1$ or $\frac{1}{2}(\vec\delta_1-\vec\delta_3)$] geometries are shown
in the second and third panels of Fig. 1(a), respectively.
}
\end{figure}

First, we find that the sliding in AB-BLG
lifts the degeneracy of optical phonon modes immediately with the splitting size
depending on both the sliding geometries
and the symmetry of the phonon modes [Fig. 2].
The degeneracy is recovered when the sliding ends up in AA-BLG [Fig. 2].
This finding is readily understood from the crystal symmetry of AB- and AA-BLG
that guarantee the degeneracy of TO and LO modes for E$_u$ and E$_g$ phonons.
However, E$_g$ and E$_u$ phonon modes exhibit completely different responses
to the sliding because of their atomic motions (one in-phase and the other out-of-phase).
The frequencies of IR-active E$_u$ phonon modes
are apparently split into two modes upon sliding
with a strong dependence on the sliding vectors [Fig. 2].
On the other hand, the frequencies of Raman-active E$_g$ modes are seemingly
unaffected at all by the sliding motion although
the exact degeneracy does not hold any more as Fig. 2 illustrates.
We observe that the splitting between TO and LO phonons
is as large as 20 cm$^{-1}$ for E$_u$ mode
but no more than 1 cm$^{-1}$ for E$_g$ mode.

To understand the intriguing behavior of optical phonons upon the sliding,
we construct a minimal model of the phonon modes in sliding BLG.
From our {\it ab initio} calculations for in-plane atomic displacements,
we find that the interlayer ion-ion interaction energy
changes very little (less than 10\%) compared with the interlayer ion-electron interaction.
So, additional force constants that account for the renormalization
of in-plane phonon frequencies by the interlayer electronic hopping are sufficient
to include into the conventional force-constant model.
Within the harmonic approximation,
the phonon frequency of BLG ($\omega_\pm$ for $E_{g(u)}$ mode)
at $\Gamma$-point can be expressed as $\omega^2_{\pm}=\omega^2_{\rm G}+\xi_\pm/m_c$, where
$\omega_{\rm G}$ is the frequency of $E_{2g}$ phonon modes of SLG
and
$\xi_\pm$ is the effective force constant
for the $E_{g(u)}$ phonon-renormalization owing to the interlayer electron hoppings in BLG
($m_c\simeq1.99\times10^{-23}$g is the mass of carbon atom)~\cite{Georgy}.
The Hamiltonian for BLG
with a sliding vector ${\vec d}=(d_x,d_y)$~\cite{son84,Falko,Gail}
can be written as
${\mathcal H}_\textrm{total}={\mathcal H}_\textrm{intra}+{\mathcal H}_\textrm{inter}$
where the intralayer Hamiltonian is
${\mathcal H}_\textrm{intra}=\sum_{\alpha}\Pi_{\vec p}c^\dagger_{B\alpha}(\vec p)c_{A\alpha}(\vec p)+\textrm{(h.c.)}$
and interlayer interaction
${\mathcal H}_\textrm{inter}=\pi_{\vec p}(\vec d)c^\dagger_{B1}(\vec p)c_{A2}(\vec p)
-\gamma_1 c^\dagger_{A1}(\vec p)c_{B2}(\vec p)+\textrm{(h.c.)}$.
Here $c_{A\alpha}$ ($c^\dagger_{B\alpha}$) is an annihilation (creation)
operator for electron at site $A\alpha$ ($B\alpha$) in the upper and lower layers ($\alpha=1,2$).
$\Pi_{\vec p}=\frac{3a_c t}{2\hbar}(p_x+ip_y)$ and
$\pi_{\vec p}(\vec d)=\frac{3a_c \gamma_3}{2\hbar}[p_x+ip_y-\beta(d_y-id_x)]$ where
$t\simeq 3.0$ eV is the nearest-neighbor ({\it nn}) intralayer-hopping constant,
$a_c$ the intralayer {\it nn} distance, $\gamma_1(\simeq t/10)$
and $\gamma_3(\simeq\gamma_1)$ the {\it nn} and the next {\it nn} interlayer-hoppings, respectively,
and $\beta\sim a_c^{-2}$~\cite{son84,Falko,Gail}.
With an intralayer electron-phonon coupling constant of $g \simeq 58$ eV/nm~\cite{Ando,Ando2},
the Hamiltonian with the phonons can be described by changing
$\Pi_{\vec p}$ to
$\Pi_{\vec p}(u)=\frac{3a_c t}{2\hbar}(p_x+ip_y)+3ug$ for the LO mode
and to $\Pi_{\vec p}(u)=\frac{3a_c t}{2\hbar}(p_x+ip_y)-3iug$ for the TO mode~\cite{Ando,Ando2}.
Here $u$ is the amplitude of $E_{2g}$ phonon modes in SLG
and then the atomic displacements at A and B sites
in the layer ${\alpha}$ are given as
${\vec u}_{A\alpha}=-{\vec u}_{B\alpha}
= u\hat{y}~(=u\hat{x})$ for the LO (TO) mode [Fig. 1(b)].
With $\gamma_{1(3)}(u)\simeq \gamma_{1(3)}+{\mathcal O}(u^2)$,
$\xi_\pm$ is obtained from the energy variation with respect to the atomic displacement,
\begin{equation}
\xi_\pm=\frac{1}{2}\frac{g_v g_s}{S_c}
\frac{\partial^2}{\partial u^2}
\int_{S_c}d{\vec p ^2}~\left[{\mathcal E}^\pm_{\vec p} (u)-{\mathcal E}^S_{\vec p}(u)\right],
\end{equation}
where
${\mathcal E}^\pm_{\vec p} (u)=-\big[\left|\Pi_{\vec p}(u)\right|^2+|\Pi_{\vec p}(\pm u)|^2+|\pi_{\vec p}(\vec d)|^2 
+\gamma_1^2+2\big|\Pi_{\vec p}(u)\Pi_{\vec p}(\pm u)+\gamma_1\pi^*_{\vec p}(\vec d)\big|\big]^{1/2}$
and ${\mathcal E}^S_{\vec p}(u)=-2\left|\Pi_{\vec p}(u)\right|$.
${\mathcal E}^\pm_{\vec p}(u)$ corresponds
to the energy of sliding BLG with $E_g$ and $E_u$ optical phonon
modes, respectively, which can be obtained by direct diagonalization of ${\mathcal H}_\textrm{total}$.
$g_v=g_s=2$ are the valley and spin degeneracy, respectively,
and $S_c=\pi p_c^2=\frac{2^2\pi^2}{3^{3/2}a_c^2}$ is half the BZ area
accounting the valley degeneracy~\cite{Georgy}.
In Eq. (1), the intralayer electron-phonon contribution [${\mathcal E}^S_{\vec p}(u)$] of two graphene layers
is subtracted from the energy of BLG with phonons [${\mathcal E}^\pm_{\vec p}(u)$] to obtain the
renormalized interlayer force constant only
since $\omega_G$ is assumed to be already renormalized intralyer phonon frequency.

The interlayer force-constant $\xi_+$ becomes negligible for $E_g$ phonon modes regardless of sliding, {\it i.e.},
$\omega_{+}\simeq \omega_{\rm G}$. This can be demonstrated easily
by shifting $\vec p$ to
$\vec p-(\frac{2\hbar ug}{a_c t},0)$ and $\vec p+(0, \frac{2\hbar ug}{a_c t})$
for LO and TO $E_g$ modes in Eq. (1), respectively.
So, this explains the reason why sliding does not change the frequency
of $E_g$ modes.
In contrast, for $E_u$ modes of AB-BLG without sliding,
we find $\xi_-\simeq\frac{3\sqrt{3} g^2\gamma_1}{\pi t^2}$ so that
$\omega_{-}(\vec d=0)-\omega_+\simeq
\frac{1}{2\omega_{\rm G}}\frac{3\sqrt{3} g^2\gamma_1}{\pi m_c t^2}\sim 10~{\rm cm}^{-1}$
in a good agreement with our calculation in Fig. 2.
With sliding along $\vec d=\alpha\vec\delta_1$ ($|\alpha|\ll 1$), we find that
$\Delta\omega_-(\vec d)\equiv\omega_-(\vec d)-\omega_-(\vec d=0)\propto
\mp\frac{\Lambda}{2\omega_G}\frac{g^2\gamma_1\gamma_3}{m_c t^3}\alpha$
for the LO (TO) $E_u$ mode (here $\Lambda$ is a dimensionless constant),
which shows the splitting of two degenerate IR-active modes upon sliding as shown in Fig. 2.
These calculations explain the different responses of Raman-active and IR-active phonon modes
in sliding BLG.

Having understood the origin of the anomalous optical phonon splittings upon sliding,
we investigate their spectroscopic consequences.
Our calculations readily indicate that the sliding systems have different
IR reflectivity and Raman spectrum when it is placed on insulating substrates
with top or bottom gates.
The single Fano-like IR spectra of gated BLG ~\cite{kuzmenko103}
will turn into the double-peak spectrum upon sliding
because of the splitting in $E_u$ phonon frequencies.
Also, its anomalous splitting in Raman G-band~\cite{Ando,Pimenta,Pinczuk}
due to mixing between $E_g$ and $E_u$ modes
will be affected by sliding.
The splitting of Raman G-band
(or spectral transfer between opposite parity optical phonon modes ) of gated BLG
is known to originate from the inversion-symmetry
breaking by the gate electric field~\cite{Ando,Pimenta,Pinczuk,Ando2}.
Therefore, we expect that the splitting in $E_u$ mode upon sliding will generate
a more complex Raman spectrum in the gated structure, {\it e.g.}, a three-peak structure.
While the gated structure is the most straightforward
way to observe the sliding-induced splitting of optical phonons,
more generic ways are desirable of detecting tiny sliding on arbitrary substrates
without the need of heavy doping using the field-effect transistor structure.
We present below conventional Raman techniques with polarizers indeed
can detect sub-Angstrom atomic misalignment by sliding.

\begingroup
\squeezetable
\begin{table}[b]
\begin{tabular}{c|c|ccc|c}
\hline
$\vec d$ & $G$  & \multicolumn{4}{c}{Raman tensors, $\mathcal{R}_\nu$ ($\nu=1,2$)} \\
\hline
\hline
0  & D$_{3d}$ & E$_g$ & $\begin{pmatrix} c&0\\0&-c\end{pmatrix}$ & $\begin{pmatrix} 0&d\\d&0\end{pmatrix}$ &
${c=-d=1.00}$\\
\hline
$-\vec\delta_1$ & D$_{6h}$ & E$_{2g}$  & $\begin{pmatrix} c&0\\0&-c\end{pmatrix}$
& $\begin{pmatrix} 0&d\\d&0\end{pmatrix}$& ${c=d=1.11}$ \\
\hline
$\frac{1}{2}\vec\delta_1$ & D$_{2h}$ &
$\begin{matrix}\textrm{A}_{g}\\\textrm{B}_{1g}\end{matrix}$ &
$\begin{pmatrix} a&0\\0&b\end{pmatrix}$ &$\begin{pmatrix} 0&d\\d&0\end{pmatrix}$
&$\begin{smallmatrix}a=+1.07, & b=-0.88, & d=-0.97 \end{smallmatrix}$\\
\hline
$-\frac{3}{4}\vec\delta_1$ & \multirow{5}{*}{C$_{2h}$}& \multirow{5}{*}{$\begin{matrix}\textrm{A}_g \\ \textrm{B}_g
\end{matrix}$} &
\multirow{5}{*}{$\begin{pmatrix} a&0\\0&b\end{pmatrix}$} &
\multirow{5}{*}{$\begin{pmatrix} 0&d\\d&0\end{pmatrix}$}
& $\begin{smallmatrix}a=+1.09,& b=-1.13,&d=-1.11 \end{smallmatrix} $\\ \cline{1-1} \cline{6-6}
$-\frac{1}{2}\vec\delta_1$ & & & & &$\begin{smallmatrix}a=+0.95,& b=-0.98,& d=-1.07 \end{smallmatrix} $\\ \cline{1-1}
\cline{6-6}
$-\frac{1}{4}\vec\delta_1$ & & & & & $\begin{smallmatrix}a=+0.91,& b=-0.90,& d=-1.00 &\end{smallmatrix} $\\
\cline{1-1} \cline{6-6}
$+\frac{1}{4}\vec\delta_1$ & & & & & $\begin{smallmatrix}a=+0.95,& b=-0.91,&d=-0.96 &\end{smallmatrix}$\\ \cline{1-1}
\cline{6-6}
$+\frac{1}{3}\vec\delta_M$ & & & & & $\begin{smallmatrix}a=+0.99,& b=-0.92,& d=-1.01 \end{smallmatrix}$\\
\hline
$+\frac{1}{6}\vec\delta_M$ & C$_{i}$ & A$_{g}$ &
$\begin{pmatrix} a&c\\c&b\end{pmatrix}$ & $\begin{pmatrix} a'&c'\\c'&b'\end{pmatrix}$
& $\begin{smallmatrix}a=-0.91, &  b=+0.88, &  c=-0.39, \\ a'=-0.33, &b'=+0.32, & c'=+0.87\end{smallmatrix}$ \\
\hline
\end{tabular}
\caption{Calculated Raman tensors for various sliding vectors ($\vec d$).
$G$ is the Schoenflies notation of the point-group symmetry for each sliding geometry.
Here, $\vec\delta_M=\vec\delta_1-\vec\delta_3$. The components of all Raman tensors
expressed in the vibrational symmetry-group notations are normalized to
AB-BLG Raman tensor.}
\end{table}
\endgroup

The position of Raman G-band of sliding BLG does not change
at all for all sliding configurations,
but the change in the crystal symmetry produces
very interesting movement of the Raman intensity.
In order to calculate the non-resonant Stocks Raman intensity of G-band,
we used the Placzek approximation for the Raman intensity ($I$) as
$I\propto\frac{d\sigma}{d\Omega}=\sum_{\nu=1,2}
|\textbf{e}_i\cdot\mathcal{R}_\nu\cdot\textbf{e}_s|^2$~\cite{bruesch,porezag,umari}.
Here $\mathcal{R}_\nu$ is the Raman tensor,
$\textbf{e}_i$ and $\textbf{e}_s$
are the polarization of the incident and scattered lights,
respectively~\cite{bruesch,porezag,umari}.
$\mathcal{R}_\nu$ associated with the doubly degenerate
Raman-active phonon modes is proportional to
the Raman susceptibility $\alpha^\nu_{ij}$, which is defined as
$\alpha^\nu_{ij}=\sum_{Ik}\frac{\partial \chi_{ij}}{\partial r_{Ik}}u^\nu_{Ik}$.
Here $u^\nu_{Ik}$ is the $\nu$th phonon eigenvector of the $I$th atom along
the $k$ direction and $\chi_{ij}$ the electric polarizability tensor
($i,j,k=x,y$, as we neglect the irrelevant $z$ component)~\cite{bruesch,porezag,umari}.
We directly calculate the derivative of $\chi_{ij}$
with respect to the atomic displacements corresponding to the $\nu$-th modes.
Assuming an incident light with an energy of 2.41 eV,
calculated Raman tensors
for various sliding geometries are summarized in Table I together with corresponding
crystal-symmetry groups and vibrational symmetries~\cite{loudon13}.
Our calculations of Raman tensors for AB-BLG
and AA-BLG reproduce the well-known results with
$E_g$ and $E_{2g}$ symmetries~\cite{loudon13,cheong2}.
Upon sliding, changes in the crystal symmetry
produce correspondingly distinctive Raman tensors.
With $\mathbf{e}_i=\mathbf{e}_s=(\cos\theta,\sin\theta)$,
the polarized Raman intensity for AB-BLG is given as
$I=|c(\cos^2\theta-\sin^2\theta)|^2+|2 c \sin\theta\cos\theta|^2
\equiv I_{AB}$
so that the intensity does not depend on the polarization angle $\theta$
[Fig. 3(a)]~\cite{loudon13,cheong2}.
This behavior is also the same for AA-BLG while with a different intensity.
When the sliding direction ($\vec {d}$)
is along $\pm\delta_1$,
the polarized Raman intensity is given
as $I(\theta)/I_{AB}=\frac{(a-b)^2}{8}\cos4\theta+\frac{a^2-b^2-d^2}{2}\cos2\theta+I_0$
($a,b$, and $d$ are given in Table I and $I_0$ is a constant).
$I(\theta)$ has an elliptic shape for $a\approx b$ or quadrupolar form otherwise
as shown in Fig. 3.
When $\vec d$ deviates from the direction of $\pm \delta_1$, the polarized intensity
tends to rotate as shown in Fig. 3(c).
In case that the overall shape of the polarized intensity looks similar,
its magnitude has a strong dependence on the sliding size [Fig.3 (b)].

\begin{figure}[t]
\includegraphics[width=1.0\columnwidth]{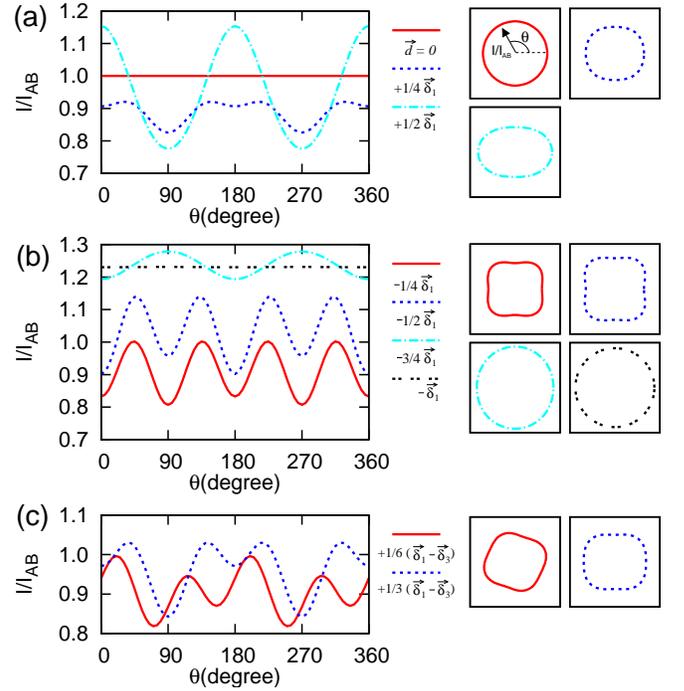}
\caption{(color online)
Polarization dependence of the Raman G-band intensity in BLG
with a sliding along (a) $+\vec{\delta}_1$,
(b) $-\vec{\delta}_1$, and (c) $+(\vec{\delta}_1-\vec{\delta}_3)$ directions.
The panels in the left show the Raman intensity normalized
to that of AB-BLG as a function of polarization angle ($\theta$)
and small rectangular pannels on the right show
corresponding polar plots.}
\end{figure}

\section{Conclusions}

In conclusion, we have shown theoretically
that tiny sliding of the layers in BLGs
can induce anomalous splitting of optical phonon
that changes the spectroscopic features significantly.
We have shown that the frequencies of the degenerate in-phase optical phonons ($E_{g}$ mode)
are hardly changed irrespective of sliding distance and direction.
On the other hand, the polarization-dependent Raman intensity associated with $E_{g}$ mode
is modified strongly
so that the sub-Angstrom misalignment between two graphene layers
can be resolved by spectroscopic methods.
We expect that this study will provide essential information
for spectroscopic measurements and for linking local atomic structures
to novel electronic properties of stacked 2D atomic crystals.

\section{Theoretical methods}

We calculated electronic structures
and phonon dispersions of the sliding BLGs
using first-principles methods
with a plane-wave basis set~\cite{pwscf}.
The local density approximation is adopted for the exchange-correlation
functional, and the phonon frequencies are calculated
using the density functional perturbation theory~\cite{dfpt}.
Computations are also repeated using the atomic-orbital basis set~\cite{siesta},
the generalized gradient approximation (GGA)~\cite{pbe},
and frozen phonon method~\cite{siesta}
to find almost identical results.
In calculating electronic structures and phonon dispersions
with the atomic orbital basis set, the basis-set superposition errors were removed
by including two ghost atoms in the unit cell~\cite{bsse}.
A semi-empirical correction of van der Waals (vdW) forces is added
to all our calculations following Grimme's proposal,
which is essential to obtain the correct interlayer distance
of sliding BLGs~\cite{grimme,son84}.
We note that our methods accurately describe the
phonon frequencies of graphitic systems
when the appropriate interlayer-distance is provided~\cite{marzari,tan}

\section{Acknowledgments}
S.-M. C. thanks C.-H. Park for discussions.
Y.-W.S. is supported by the NRF of Korea grant funded by MEST
(QMMRC, No. R11-2008-053-01002-0 and Nano R\&D program 2008-03670). 
S.-H.J. acknowledges the support from the NRF of Korea
grant funded by MEST (SRC program No. 2011-0030046 
and WCU program No. R31-2008-000-10059).
Computational resources have been provided 
by KISTI Supercomputing Center (Project No. KSC-2011-C1-21) and the CAC of KIAS.

\end{document}